\documentclass[12pt]{article}
\usepackage{geometry} % see geometry.pdf on how to lay out the page. There's lots.
\usepackage{textcomp} % provide lots of new symbols
\usepackage{graphicx}  % Add graphics capabilities
\usepackage{flafter}  % Don't place floats before their definition
\usepackage{amsmath,amssymb}  % Better maths support & more symbols
\usepackage{bm}  % Define \bm{} to use bold math fonts

\usepackage[pdftex,bookmarks,colorlinks,breaklinks]{hyperref}  
\hypersetup{linkcolor=black,citecolor=black,filecolor=black,urlcolor=black}
\usepackage{memhfixc}  
\usepackage[english]{babel}
\usepackage[T1]{fontenc}
\usepackage[ansinew]{inputenc}

\begin{document}

\title{Linear Two-Dimensional MHD of Accretion Disks:
Crystalline structure and Nernst coefficient}

\author{Giovanni Montani$^{1,2,3,4\star}$ and Riccardo Benini$^{3,4,\dag}$\\
\footnotesize{$^{1}$ ENEA - C.R. Frascati (Department F.P.N.)}, 
\footnotesize{Via Enrico Fermi, 45 (00044), Frascati (Rome), Italy}\\
\footnotesize{$^{2}$ ICRANet - C.C. Pescara, P. della Repubblica, 10 (65100), Pescara, Italy}\\
\footnotesize{$^{3}$Department of Physics (G9) - ``Sapienza'' Universit\`a di Roma,}
\footnotesize{Piazzale A. Moro, 5 (00185), Rome, Italy}\\
\footnotesize{$^{4}$ ICRA - International Centre for Relativistic Astrophysics}\\
$^{\star}$ montani@icra.it\\
$^{\dag}$ riccardo.benini@icra.it}

\date{September $2^{nd}$, 2009}

\maketitle
\begin{abstract}
We analyse the two-dimensional MHD configurations characterising
the steady state of the accretion disk on a highly
magnetised neutron star.
The model we describe has a local character and represents
the extension of the crystalline  structure
outlined in \cite{C05}, dealing with a local model too,
when a specific accretion rate is taken into account.
We limit our attention to the linearised MHD formulation
of the electromagnetic back-reaction characterising the 
equilibrium, by fixing the structure of the
radial, vertical and azimuthal profiles.
Since we deal with toroidal currents only, the consistency
of the model is ensured by the presence of a small
collisional effect, phenomenologically described by
a non-zero constant Nernst coefficient
(thermal power of the plasma).
Such an effect provides a proper balance of the electron
force equation via non zero temperature gradients,
related directly to the radial and vertical velocity
components.

We show that the obtained profile has the typical
oscillating feature of the crystalline structure,
reconciled with the presence of viscosity, associated
to the differential rotation of the disk, and with
a net accretion rate. In fact, we provide a direct relation between the
electromagnetic reaction of the disk and the
(no longer zero)
increasing of its mass per unit time.
The radial accretion component of the velocity results to be
few orders of magnitude below the equatorial sound velocity.
Its oscillating-like character does not allow a real
matter in-fall
to the central object
(an effect to be searched into non-linear MHD corrections),
but it accounts for the out-coming of steady fluxes, favourable to the
ring-like morphology of the disk.

\textbf{keywords:} \textsf{Accretion disks; Plasma physics; Nernst coefficient}

\textbf{PACS Nos.:} \textsf{97.10.Gz, 52.30.-q}

\end{abstract}

\section{Introduction}

A long standing problem in astrophysics concerns
the mechanism of accretion that compact objects manifest
in the presence of lower dense companions.
It is  well established by observations\cite{verbunt82}
that such an accretion profile takes place often
with the morphology of a thin disk configuration.
Since from the very beginning of these studies,
increasing interest raised about the details of
the angular momentum transport across the disk
structure. The solution of this problem appears
settled down in the case of a compact object, for
which the intrinsic magnetic field does not
exert a significant Lorenz force on the elementary
charged constituents of the in-falling plasma.
In this limit, the fluid-dynamics
description appears well-grounded and the resulting
axisymmetric configurations well understood (see among the first analyses of the problem\cite{S73,pringlerees72,lyndenpringle74}; for a relativistic analysis of plasma accreting into a black hole, see \cite{rw}).
In the fluid-dynamics paradigm, the accretion mechanism
relies on an angular momentum transfer which is
allowed by the shear viscosity properties of the
disk material. The differential angular rotation
of different disk layers is associated with a
non-zero viscosity coefficient, which accounts 
for the diffusion and turbulence phenomena
emerging in the micro-scale structure of the fluid.

However, when the magnetic field of the central object
is sufficiently high, the electromagnetic
back-reaction
of the disk plasma becomes very important.
As shown in \cite{C05}, the Lorenz force introduces
a crucial coupling between the radial and the vertical
equilibrium, which deeply alters the morphology
of the system. In particular, the radial dependence of
all the fundamental quantities acquires
an oscillating
character, modulating the background profile. Furthermore, as outlined in \cite{CR06}, when
the disk plasma is characterised by a $\beta$-parameter
close to the unity, such an oscillating-like
behaviour stands as the dominant effect and the
disk is indeed decomposed in a ring profile.
However, such a two-MHD approach is pursued
neglecting the accretion rate of the disk in the leading
order and thus avoiding the problem
concerning the azimuthal balance of the force acting
on the electrons. In fact, in these two works,
toroidal currents and matter fluxes are addressed only.

This question about a consistent electron force balance
along the azimuthal direction (the azimuthal electric field
is request to vanish by the axial symmetry) affects also
those works where the $z$-dependence of the system is
averaged out for a thin disk\cite{B01}. Thus,
the possibility of presenting a self-consistent MHD
description for an accreting disk configuration,
calls  significant attention, especially in view of the
implication that the disk morphology can have on the
formation of jets from compact astrophysical structures\cite{lynden-bell}.
For a discussion which properly traces the way to
face this relevant configuration problem, see \cite{cproceeding}.

This paper presents a solution to the two-dimensional
MHD structure of a thin accretion disk, in the
linear regime, when the role of a Nernst coefficient
(the so-called plasma thermal power) is taken into account.
In particular, we are reconciling the oscillating
behaviour outlined by B. Coppi, with the accretion
features of the disk. In fact, in our model, radial
and vertical velocities are included in the problem,
still retaining the poloidal character of the
currents living in the plasma. The resulting equilibrium
configuration is associated with a radial velocity which,
following the poloidal currents, oscillates as a
function of the radial coordinates and decays in the
vertical direction. Despite its oscillating-like character,
this local radial velocity is responsible for a non-zero
accretion rate and it describes a mechanism for the
appearance of a steady matter flux, able to enforce
the ring-profile associated with the crystalline structure.
Nonetheless, the oscillating character 
of this radial velocity prevents a real matter in-fall
toward the centre of symmetry.
Thus, a real accretion of the central object seems to
require, in this scheme, the presence of non-linear MHD features;
but the local character of the model
does not allow to properly characterise the global accretion profile
of the central object, that could imply
a connection with the boundary layer physics\cite{SS88}.

The structure of the paper is as follows.
In Section \ref{sec2} we present a description of the main
features characterising an accretion disk model.
The tools needed to construct the configuration scheme
are provided. Section \ref{sec3} is devoted to fix the local
nature of the model, properly defining the addressed
approximations. In Section \ref{sec4}, we build up the
system of partial differential equations which
governs the two-dimensional MHD equilibrium,
by assigning the radial, vertical and azimuthal force balance.
Section \ref{sec5} is concerned with the analysis of the linearised
equations, based on the assumption that the magnetic
field induced into the plasma, by the presence of the
toroidal currents is much less in strength than the external
field, given by the central object.
Section \ref{sec6} faces the discussion of the electron force balance
equation, by introducing a non-zero and constant Nernst
coefficient to preserve the presence of radial and vertical
non-vanishing components of the matter velocity.
Finally we develop some
phenomenological consideration on the model,
mainly aimed to estimate the strength of the radial
velocity and the associated accretion rate.
Furthermore, we fix the condition to neglect the temperature
gradients into the radial equilibrium.
Brief concluding remarks follow in Section \ref{secConclusioni}.

\section{Model for an Accretion Disk \label{sec2}}

Let us fix here the basic statements concerning
the steady MHD regime for the
specific case of an accretion disk configuration
around a compact astrophysical object 
(a mass over the critical value to have a neutron
star within a radius of
few kilometres), which is also 
strongly magnetised (a dipole-like field of
about $10^{12} Gauss$). Thus, we deal with
a typical pulsar, accreting material from
a binary companion via a thin disk structure
made up of rotating plasma in equilibrium.

To adapt the MHD equations to the disk symmetry, we
introduce cylindrical coordinates $\{ r\, ,\phi\, ,z\}$.
The gravitational potential of the pulsar takes, in this
coordinates, the form
\begin{equation}
\chi (r\, ,z) = -\frac{GM}{\sqrt{ r^2 + z^2}}
\, ,
\label{Gravpot}
\end{equation}
where $G$ denotes the Newton's constant and $M$
the mass of the spherical central object,
while the self-gravitation of the disk is regarded
as a negligible effect.

To describe the magnetic field, we take the
potential vector in the form
\begin{equation}
\vec{A} = \partial _r\Pi \vec{e}_r + \psi \vec{e}_{\phi}
+ \partial _z\Pi \vec{e}_z
\, ,
\label{potvec}
\end{equation}
where $\Pi (r\, , z)$ and $\psi (r\, ,z^2)$ are
arbitrary functions, but only the last one is a
physical degree of freedom, since the magnetic field
reads\footnote{A $B_{\phi}$-component of the magnetic field
can also be included in the theory by a generic
term $K\vec{e}_{\phi}$, taken, for instance,
as a function of $\psi$ (i.e. $K=K(\psi )$).
%in order to preserve the corotation theorem, like in Ferraro (1937)
%\cite{F37}.
%This requirement is equivalent to deal,
%in the disk, with toroidal (azimuthal) current only.
For the sake of simplicity and in agreement with
Coppi (2005)\cite{C05} and 
Coppi-Rousseau (2006)\cite{CR06}, 
in our treatment we will fix this component
equal to zero.}
\begin{equation}
\vec{B} = -\frac{1}{r}\partial _z\psi \vec{e}_r +
\frac{1}{r}\partial _r\psi \vec{e}_z  
\, .
\label{vectorb}
\end{equation}
The flux function has to be decomposed as
$\psi = \psi _0 + \psi _1$, where $\psi _0$
accounts for the dipole-like magnetic field of the pulsar
(i.e. $\psi _0 = \mu r^2(r^2 + z^2)^{-3/2}$, with
$\mu = const.$),
while $\psi _1$ is a contribution due to the
toroidal currents rising in the disk configuration.
Recalling that the axial
symmetry prevents any dependence on the
azimuthal angle $\phi$ of all the quantities involved in
the problem, the continuity equation,
associated to the mass density $\epsilon$ and to the
velocity field $\vec{v}$, 
takes the explicit form
\begin{equation}
\frac{1}{r}\partial _r\left(r\epsilon v_r\right) +
\partial _z\left(\epsilon v_z\right) = 0
\, ,
\label{conteq}
\end{equation}
which admits the solution
\begin{equation}
\epsilon \vec{v} =
-\frac{1}{r}\partial _z\Theta \vec{e}_r +
\epsilon \omega (r\, ,z)r\vec{e}_{\phi } +
\frac{1}{r}\partial _r\Theta \vec{e}_z
\, ,
\label{solconteq}
\end{equation}
$\Theta = \Theta (r\, ,z)$ being a generic
function, but for its odd symmetry
in the $z$-coordinate.
This property has to be required
in order to ensure
that the accretion rate of the disk,
as averaged over
the vertical direction, be non-vanishing, i.e.
\begin{equation}
\dot{M}_d = -2\pi r\int _{-z_0}^{z_0}\epsilon v_rdz =
4\pi \Theta (r\, , z_0) \equiv 2\pi I 
\neq 0
\, ,
\label{Mdot}
\end{equation}
where $z_0$ is the half depth of the disk
and we recall that $v_r < 0$ on the vertical average,
to ensure a real accretion of the disk.

The different symmetry of the functions $\psi$
and $\Theta$ with respect to the $z$-dependence,
prevents to fix \emph{a priori} a relation between
them. The angular velocity $\omega (r\, ,z^2)$,
describing the differential rotation of the disk,
is an even function of $z$ and, by virtue of
the corotation theorem\cite{F37}, can be taken as
a function of the flux surfaces, i.e.
$\omega = \omega (\psi )$.

The momentum transfer through the disk structure
is characterised by an azimuthal friction between
the radial layers, properly described by a viscosity
coefficient $\eta$, responsible for the corresponding
non-vanishing stress tensor components.
The coefficient $\eta$ can be decomposed as
$\eta = \eta _0 + \eta _1$, where the $\eta _1$
component is associated with the additional effects
arising from the toroidal currents.
The necessity to include
the electromagnetic back-reaction terms 
in the viscosity coefficient,
is a consequence of the specific coupling
such back-reaction establishes. In fact,
it was demonstrated\cite{C05} that the currents
induced in the plasma of the disk, provide an intrinsic
local coupling between the radial and the vertical
equilibrium, absent in the lowest-order approximation.

\section{Local configuration of the thin disk \label{sec3}}

Let us now specialise our model to the case of a
thin disk, i.e. $z_0/r\ll 1$ over the whole
configuration.
On average, 
the dominant contribution to the angular
velocity of the plasma particles, is then
the (equatorial) Keplerian value
$\omega _K \equiv (GM/r^3)^{\frac{1}{2}}$.

For a thin disk, the
$z$-component of the velocity can be
properly identified
(see Shakura (1973) \cite{S73})
with the sound velocity
$v_s \equiv \sqrt{2K_BT/m_i}$
($K_B$ denoting the Boltzmann constant,
$T$ the plasma temperature,
and $m_i$ the ion mass), entering
the fundamental inequality
\begin{equation}
\frac{z_0}{r} \sim \frac{v_s}{v_{\phi}}\ll 1
\, ,
\label{thincond}
\end{equation}
which ensures the possibility to neglect the
vertical motion with respect to the azimuthal one.

To estimate on average the ratio between the radial
velocity and the azimuthal one, we can follow the
simple equilibrium condition to fix the asymptotic
radial velocity of a plasma element falling
into the disk
\begin{equation}
\nu _cv_r \sim \omega ^2r = \omega v_{\phi}
\quad \Rightarrow \quad
\frac{v_r}{v_{\phi }} \sim \frac{\omega }{\nu _c}
\, ,
\label{eqeq}
\end{equation}
$\nu _c$ being a characteristic frequency of
particle collision. A rough estimation provides
$\nu_c \sim v_sn^{1/3}$, where $n$ denotes the
number density. Taking $v_s \sim 10^{-3}c$ and
$n\sim 10^{10}cm^{-3}$, we get
$\nu _c \sim 10^{9}Hz$, a value much greater than
the typical Keplerian frequency.
Thus, we conclude that $v_{\phi} \sim \omega _Kr$
is, in practice, responsible for the main matter
flow within the disk, despite its stationary accretion takes
place along the radial direction.

In this limit of approximation, the MHD condition
on the balance of the Lorenz force has 
dominant radial and vertical components, 
providing the electric field in the form
expected by the corotation theorem, i.e.
\begin{equation}
\vec{E} = -\frac{\vec{v}}{c}\wedge \vec{B} =
-\frac{d\Phi}{d\psi}\vec{\nabla }\psi =
-\frac{\omega }{c}\left( \partial _r\psi \vec{e}_r
+ \partial _z\psi \vec{e}_z\right)
\, ,
\label{MHDeqcond}
\end{equation}
$\Phi$ denoting the electrostatic potential.
However, the axial symmetry prevents a dependence
of $\Phi$ on the azimuthal angle and hence, the
corresponding $\phi$-component of the electric field
vanishes identically. This fact requires an additional
effect to be included into the problem, able to
balance the non-vanishing Lorenz force in the
azimuthal direction, due to the velocity components
$v_r$ and $v_z$.
We will address this crucial question in
Section 6, in the limit of a linear theory.

In the thin disk approximation, the strength of the
gravitational field $\vec{G}$ reads as
\begin{equation}
\vec{G} = -\omega _K^2r\vec{e}_r -\omega _K^2z\vec{e}_z
\, .
\label{gravfield}
\end{equation}
We see that a thin disk configuration is justified
only in the presence of a sufficiently high rotation
to deal with a confining vertical gravitational force
(for the linear scheme, here addressed, the Lorenz force
can not affect this statement\cite{CR06}).

According to the analysis in \cite{C05}, we now
develop the disk configuration around a given value
of the radial coordinate $r_0$, limiting our attention
to a narrow enough interval to express the local
angular velocity field in the form
\begin{equation}
\omega \simeq \omega_K + \delta \omega \simeq
\omega _K + \frac{d\omega }{d\psi _0}\psi _1
\equiv \omega _K + \omega ^{\prime}_0\psi _1
\, ,
\label{svilomeg}
\end{equation}
where $\psi _1 = \psi _1(r_0\, ,z^2\, ,r-r_0)$.
We approximate the dipole surface function, at $r_0$,
as $\psi _0(r_0)\simeq \frac{\mu }{r_0}$. We drop the
relic $z$-dependence in view of the
thin nature of the disk.

In what follows, the analysis is performed by
addressing the drift ordering approximation that
fixes the dominant character of the second spatial
derivatives of $\psi_1$.
%Finally, it is worth noting that the scheme presented
%above allows to separate the dynamics at $r_0$,
%which will resembles the one proposed in \cite{S73}  
%(see also Bisnovatyi-Kogan (2001) \cite{B01}),
%from the effects proper of plasma physics.
In particular, the profile of the disk, so outlined, will include
viscous features, as in \cite{S73}, but making
account for the crystalline structure derived in \cite{C05}.  
Indeed, reconciling the accretion feature of the
disk with the plasma effects, as described in a
bidimensional MHD approach, we will need a significant
deviation from the standard mechanism of angular momentum
transfer, in the sense that the azimuthal equation holds 
at the level of electromagnetic
back-reaction only.

\section{Configuration Equations \label{sec4}}

In order to fix the profile of the disk around the
configuration at $r_0$, we have to provide the equations
governing the radial, the vertical and the tangential
equilibrium in the
presence of the toroidal currents.
The radial and the vertical configurations are not significantly
affected by the viscosity, since
its presence mainly concerns  the differential rotation of
the disk\cite{S73}. Therefore, these systems stand in the same form
as in \cite{C05}, while the implications due to a non-zero
viscous stress are summarised by the azimuthal equilibrium.

In order to cast the whole system, we split the energy density
and the pressure in the form
\begin{equation}
\epsilon = \bar{\epsilon}(r_0, \, z^2) +
\hat{\epsilon} (r_0,\, z^2,\, r-r_0)
\quad , \quad
p = \bar{p}(r_0, \, z^2) +
\hat{p}(r_0,\, z^2,\, r-r_0)
\, ,
\label{split}
\end{equation}
where the barred quantities are the contributions
existing in absence of the toroidal currents, while
the terms denoted with a hat are induced by such an 
electromagnetic reaction. According to the framework we outlined, the 
vertical equilibrium is governed by the two
relations
\begin{eqnarray}
\label{verticalequilibrium}
D(z^2) \equiv \frac{\bar{\epsilon}}{\epsilon _0(r_0)} =
\exp ^{-\frac{z^2}{H_0^2}}
\, , \, \epsilon _0(r_0) \equiv \epsilon (r_0,\, 0)
\, , \, H_0^2 \equiv \frac{2 K_B\bar{T}}{m_i\omega _K^2}\;,\\
\partial _z\hat{p} + \omega ^2_Kz\hat{\epsilon}
+ \frac{1}{4\pi r_0^2}\left(
\partial ^2_z \psi_1 + \partial^2_r\psi _1\right)
\partial _z\psi_1 = 0 
\end{eqnarray}
The radial equation underlying the equilibrium
of the rotating layers of the disk, takes the form
\begin{eqnarray}\label{radialequilibrium}
\omega \simeq \omega _K + \delta \omega \simeq
\omega _0(\psi _0) + \frac{d\omega _0}{d\psi _0}\psi _1\;,\\
2\omega _Kr_0(\bar{\epsilon} + \hat{\epsilon})
\frac{d\omega _0}{d\psi _0}\psi _1 -
\frac{1}{4\pi r_0^2}\left(
\partial ^2_z \psi_1 + \partial^2_r\psi _1\right)
\partial _r\psi_1 = \nonumber\\
=\partial _r\left[
\hat{p} + \frac{1}{8\pi r_0^2}
\left(\partial_r\psi_1\right)^2\right]
+ \frac{1}{4\pi r_0^2}\partial_r\psi _1 \partial^2_z\psi_1
\end{eqnarray}
The azimuthal equation exactly reads
\begin{equation}
\epsilon v_r\partial _r(\omega r) +
\epsilon v_z\partial _z(\omega r) +
\epsilon \omega v_r = \frac{1}{r^2}\partial _r\left(
\eta r^3\partial _r\omega \right) +
\partial _z\left[ \eta \partial _z(\omega r) \right] 
\, .
\label{exazeq}
\end{equation}
By using equation (\ref{solconteq}),
the tangential equilibrium above easily
stands as 
\begin{equation}
-\partial _z\Theta \partial _r\omega
+\partial _r\Theta \partial _z\omega
-\frac{2}{r}\partial _z\Theta \omega =
3\eta \partial _r\omega + r\left[
\partial _r\eta \partial _r\omega +
\partial _z\eta \partial _z\omega +
\eta \left(\partial _r^2\omega + \partial _z^2\omega
\right) \right]
\, .
\label{rewtaneq}
\end{equation}
Accounting for the corotation theorem
$\omega = \omega (\psi)$ we can restate the spatial
derivatives of $\omega$, in terms of the corresponding
ones taken on $\psi$.
Recalling the expression
$\eta = \eta _0 + \eta _1$,
we can now make the reasonable assumption
that the viscosity correction $\eta _1$ be
written as
\begin{equation}
\eta _1(r_0\, , \psi _1 ) \sim
\left( \frac{d\eta _1}{d\psi _1}
\right) _{\psi _1= 0\, ,r=r_0} \psi _1
\equiv \eta ^{\prime }_0\psi _1
\, ,
\label{formeta_1}
\end{equation}
we recall that for vanishing $\psi _1$,
the correction $\eta _1$ vanishes too.
In the work \cite{S73}, a valuable proposal for the
expression describing the viscosity coefficient
$\eta _0(r_0)$ is provided, i.e.
\begin{equation}
\eta _0(r_0) \equiv \frac{2}{3} \alpha \epsilon _0
{v_s}_0z_0
\, ,
\label{formeta}
\end{equation}
where ${v_s}_0$ denotes the sound velocity
on the equatorial plane and $\alpha$ is
a parameter, whose value must be assigned.
Putting together all these statements,
the tangential equation rewrites
\begin{eqnarray}
\label{azeqred}
-\partial _z\Theta \left(\partial _{r_0}\psi _0 +
\partial _r\psi _1\right) 
+\partial _r\Theta \partial _z\psi _1 
-\frac{2}{r_0}\partial _z\Theta
\frac{\omega _0}{\omega ^{\prime }} =
\nonumber \\
r_0\left[ \partial _r\psi _1
\eta ^{\prime }_0 \partial _{r_0}\psi _0 +
\partial _{r_0}\eta _0\partial _r\psi _1 +
\eta ^{\prime }_0
\left( \partial _r\psi _1\right) ^2 +
\eta ^{\prime }_0
\left( \partial _z\psi _1\right) ^2 + \eta _0(r_0)
\left( \partial ^2_r\psi _1 +
\partial ^2_z\psi _1\right) \right]
\, .
\end{eqnarray}
Since we have
$\partial _{r_0}\eta _0\sim \eta _0/r_0$,
the first term on the right-hand-side is negligible
in comparison to the term containing
second derivatives. Furthermore, the request
$\eta _1\ll \eta _0$
implies that
$\eta ^{\prime }_0 \ll \eta _0/\psi _1$,
ensuring that also the quadratic terms in the
derivatives of
$\psi _1$
are much smaller than the remaining ones.

Therefore the form taken by the azimuthal equation for the
steady state of the disk reads as
\begin{eqnarray}
\label{azeqfin}
-\partial _z\Theta \left(\partial _{r_0}\psi _0 +
\partial _r\psi _1\right) 
+\partial _r\Theta \partial _z\psi _1 
-\frac{2}{r_0}\partial _z\Theta \frac{\omega _0}
{\omega ^{\prime }_0} =
\nonumber \\
r_0\left[ \partial _r\psi _1
\eta ^{\prime }_0 \partial _{r_0}\psi _0 +
\eta _0(r_0)
\left( \partial ^2_r\psi _1
+ \partial ^2_z\psi _1\right) \right]
\, .
\end{eqnarray}
This equation provides a differential relation between
the flux surface $\psi _1$ and the function $\Theta $
characterising the radial and vertical matter fluxes.

\section{Configuration in the linear approximation \label{sec5}}

Let us analyse the compatibility between the radial and
the azimuthal equations, in the limit when the
induced $z$-field $B_z^1$ is much smaller than the
source one $B_{z0}$, i.e.
\begin{equation}
\frac{B^1_z}{B_{0z}} \sim k_0r_0\frac{\psi _1}{\psi _0}
\ll 1
\, , 
\label{lincon}
\end{equation}
where the magnetic surface $\psi _1$ admits the
explicit dependence
\begin{equation}
\psi _1 = \psi _1 \left(
r_0,\, k_0(r - r_0),\, \frac{z^2}{\Delta ^2}\right)
\, .
\label{ffpsi}
\end{equation}
Here, we set 
$k_0^2 = 3\omega ^2_K/v_{A0}^2$, with
$v_{A0}^2 = B^2_{z0}/4\pi \epsilon _0$
being the Alfv\'en velocity in the plasma.
Furthermore, 
$\Delta $ denotes a narrow interval for the
localisation of the $z$-dependence.
Under these hypotheses, the radial equation rewrites
\begin{equation}
\partial ^2_r\psi _1 + \partial ^2_z\psi _1 =
-k_0^2D(z^2)\psi _1
\, .
\label{ffpsix}
\end{equation}
At lowest orders in $z$, we take the following
expansion
$D(z^2)\simeq 1 - z^2/H_n^2$, $H_n$ being of
the same order of magnitude of $H_0$.
In this limit, the radial equation admits
the solution discussed in \cite{C05} which
oscillates as a sin function along the radial
direction, while it exponentially decays in $z^2$
along the vertical configuration, i. e.
$\psi _1$ reads as
\begin{equation}
\psi _1 = \psi _1^0\sin \left[ k(r - r_0)\right]
\exp \{ -\frac{k_0z^2}{H_n}\}
\, ,
\label{sollineq}
\end{equation}
where we set $k\equiv k_0\sqrt{1-\frac{1}{k_0H_n}}$
and we obtained the identification
$\Delta ^2 \equiv H_n/k_0$.
Such a structure is the linear feature of the
crystalline morphology induced in the disk by the
toroidal currents.

In the approximation where $k_0r_0\gg 1$,
we can require that the contribution due to the
currents on the viscosity coefficient be sufficiently
small to neglect the term in $\eta ^{\prime }_0$ of
 (\ref{azeqfin}), i. e. 
by virtue of the inequality
\begin{equation}
\frac{\eta ^{\prime }_0\psi _0}{\eta _0} \ll k_0r_0
\, .
\label{ineqvis}\end{equation}
Thus, the final form we address for the azimuthal equation is
as follows 
\begin{equation}
-\partial _z\Theta \left(\partial _{r_0}\psi _0 +
\frac{2\omega _0}{r_0\omega ^{\prime }_0} + 
\partial _r\psi _1\right)
+\partial _r\Theta \partial _z\psi _1 = 
r_0\eta _0(r_0)
\left( \partial ^2_r\psi _1
+ \partial ^2_z\psi _1\right) 
\, .
\label{azeqfinal}
\end{equation}
We use this equation to complete the configuration
scheme of our thin disk equilibrium. Under the same conditions,
fixed above for the linear regime,
the azimuthal equation (\ref{azeqfin}) is restated as follows
\begin{equation}\label{azeqfinlin}
-\partial _z\Theta \left(\partial _{r_0}\psi _0 +
\frac{2\omega_0}{r_0\omega ^{\prime }_0}\right)
+\partial _r\Theta \partial _z\psi _1 = \eta _0r_0\left( \partial ^2_r\psi _1
+ \partial ^2_z\psi _1\right) 
\, .
\end{equation}
Let us now assume that the term containing the
derivative $\partial _r\Theta$ be negligible
(this is natural because it is multiplied by the
small $r$-component of the magnetic field)
and then divide the equation by
$\partial _{r_0}\psi _0$.
Observing that
$\omega ^{\prime }_0\partial _{r_0}\psi _0
= \partial _{r_0}\omega _0 = -3\omega _0/2$ (where
$\omega _0 \equiv \omega_{K}$), we rewrite (\ref{azeqfinlin})
as follows
\begin{equation}
\frac{1}{3}\partial _z\Theta =
\frac{\eta _0r_0}{\partial _{r_0}\psi _0}
\left( \partial ^2_r\psi _1 + \partial _z^2\psi _1\right)
\, .
\label{rfq}
\end{equation}
Comparing the equation above with
(\ref{ffpsix}), we arrive to the relation
\begin{equation}
\epsilon _0(r_0)v_r = -\frac{1}{r_0}\partial _z\Theta =
3\eta _0k_0Y                  
\, , 
\label{v-r}
\end{equation}
in which we made use of the definition
of the dimensionless function
$Y \equiv k_0\psi _1/\partial _{r_0}\psi _0$.
Comparing the two equations, we neglected the factor
$D(z^2)$ with respect to the $z$-dependence of
$\psi _1$, because $H_n\gg \delta$, i.e.
$k_0H_n\gg 1$.

Such an accreting profile provides a net
increasing of the disk mass, according to the relation
\begin{equation}
\dot{M}_d =
-2\pi r\int _{-\infty}^{\infty}\epsilon v_rdz \simeq 
6\sqrt{2}\pi ^{3/2}\eta _0(k_0r_0)^2
\frac{\Delta \psi _1^0}{\mu }r\sin k(r - r_0)
\, , 
\label{Mdotav1}
\end{equation}
where we made use of the dipole-like
expression for $\psi _0$ and we extended the integration
over the vertical coordinate up to infinity.

It is worth noting that,
despite the radial velocity oscillates, the radial matter flux is,
on average, an in-falling one. In fact, averaging the
expression above between two nodes around $r_0$,
we get
\begin{equation}
\langle \dot{M}_d\rangle =
12\sqrt{2}\pi ^{5/2}\eta _0(k_0r_0)^2
\frac{\Delta \psi _1^0}{\mu k^2}
\, . 
\label{Mdotav}
\end{equation}
Equation (\ref{v-r}) can be easily solved for
$\Theta$ in agreement with the adopted approximation
of neglecting the term containing
$\partial _r\Theta$, as follows
\begin{equation}
\Theta =
-3\eta _0 k_0 r_0\int Ydz               
\, . 
\label{rael}
\end{equation}
Hence we get the vertical velocity in the form
\begin{equation}
v_z = \frac{1}{\epsilon_{0}r_0}\partial _r\Theta =
-\frac{3\eta _0k_0k}{\epsilon _0}\int Ydz
\, . 
\label{v-z}
\end{equation}
We see that, in the linear approximation, the azimuthal
equation ensures the existence of a non-zero accretion
rate, which is modulated by the oscillating profile of
the disk configuration.

\section{The Electron Force Balance Equation \label{sec6}}

Despite we have properly argued that the radial and
vertical components of the fluid velocity are significantly smaller
than the $\phi$-component of the velocity,
nevertheless, the MHD electron force balance,
restated here
\begin{equation}
\vec{E} + \frac{\vec{v}}{c}\wedge \vec{B} = 0
\, , 
\label{efb}
\end{equation}
requires $(\vec{v}\wedge \vec{B})_{\phi } = 0$,
because of $E_{\phi }\equiv 0$,
by virtue of the axial symmetry.
For instance, this feature arises naturally if the
corotation theorem is addressed.
More simply, this request comes out from the
independence of $\phi$ characterising the electrostatic
potential.
Thus, on the left-hand side of equation (\ref{efb}) an
additional term has to appear in order to save the
model consistence.
In the limit of the linear theory, 
the most natural term to be included in the analysis,
preserving the MHD scenario we addressed, is
a weak collisional effect, as phenomenologically
described by a non-zero constant 
Nernst coefficient. However, to introduce this hypothesis we need to give up
the idea, pursued above, that the plasma in the disk be
perfectly isothermal. Thus, we now introduce temperature 
gradients\cite{c08} in the spirit that they do not affect
significantly the three configuration equations
developed in the previous sections.
We discuss only the implications
for the electron force balance when a Nernst
coefficient is present and then we fix the restriction
on the model parameters to get full consistence.

According to this idea, 
we split the temperature as
$T=T_0 + T_1(r,\, z^2)$, with $T_0=const.$ and
$T_1\ll T_0$.
In the presence of a non-zero Nernst coefficient
$\mathcal{N}$,
the equation (\ref{efb}) takes the form
\begin{equation}
\vec{E} + \frac{\vec{v}}{c}\wedge \vec{B} =
\mathcal{N}\vec{B}\wedge \vec{\nabla }T
\, .
\label{efb3}
\end{equation}
Assuming that the gradients of the temperature are negligible
in the radial and vertical component of this equation,
its $\phi$-component provides
\begin{equation}
v_r = -\mathcal{N}c\partial _rT_1
\quad , \quad
v_z = -\mathcal{N}c\partial _zT_1
\, .
\label{solefb}
\end{equation}
We now have to observe that the temperature of the plasma
in the static MHD framework, must obey the
equation
\begin{equation}
\vec{v}\cdot \vec{\nabla }T +
\frac{2}{3}T\vec{\nabla }\cdot \vec{v} = 0
\, .
\label{temeq}
\end{equation}
Expressing the gradients of the temperature
by the relation (\ref{solefb}), the equation above
takes the form
\begin{equation}
\frac{1}{r}\partial _r(rv_r) + \partial _zv_z 
= \frac{3}{2\mathcal{N}T_0c}\left(
v_r^2 + v_z^2\right) 
\, .
\label{temeqre}
\end{equation}
It is easy to see that this equation is automatically
satisfied in the linear regime, by virtue
of the expressions for the radial and vertical
velocity (\ref{v-r}) and (\ref{v-z}) respectively.

The complete consistence of our linear model is then
guaranteed by requiring that the radial temperature
gradient be negligible in the corresponding configuration
equation. Such gradients arise in the radial equation
from the spatial variation of the pressure
$p = 2 \epsilon K_BT/m_i$
(from which we get
$\hat{p} = 2 K_BT_0\hat{\epsilon}/m_i +
2 \epsilon _0K_BT_1/m_i)$. 
The condition for the negligibility
of this term, can be stated, by virtue of
equations (\ref{v-r}) and (\ref{solefb}),
in the form
\begin{equation}
\partial _{r_0}\psi _0\gg r_0\sqrt{\frac{6\eta _0
K_B}{m_i\mathcal{N}c}}
\, , 
\label{temeqre1}
\end{equation}
To get this result, we made account for the fact that
obtaining equation (\ref{ffpsix}), we divided the
original radial equation by the term
$\partial _{r_0}\psi _0/r_0^2$. 
By making use of the expression (\ref{formeta})
for $\eta _0$,
we can rewrite the condition above in terms of the
Alfv\'en and equatorial sound velocities as
$v_A\gg \delta {v_s}_0$, with
$\delta \equiv
\sqrt{\frac{\alpha  H_0 
{v_s}_0}{4\pi \mathcal{N}T_0c}}$
(we approximated everywhere the sound velocity with its value
on the equatorial plane and we have taken $z_0\equiv H_0$).
Furthermore we implicitly
required that the radial variation of
$\hat{\epsilon}$
be smaller or, at most, equal to the 
term with the temperature gradient.

\begin{figure}[ht]
   \centering
   \includegraphics[width=.6\textwidth]{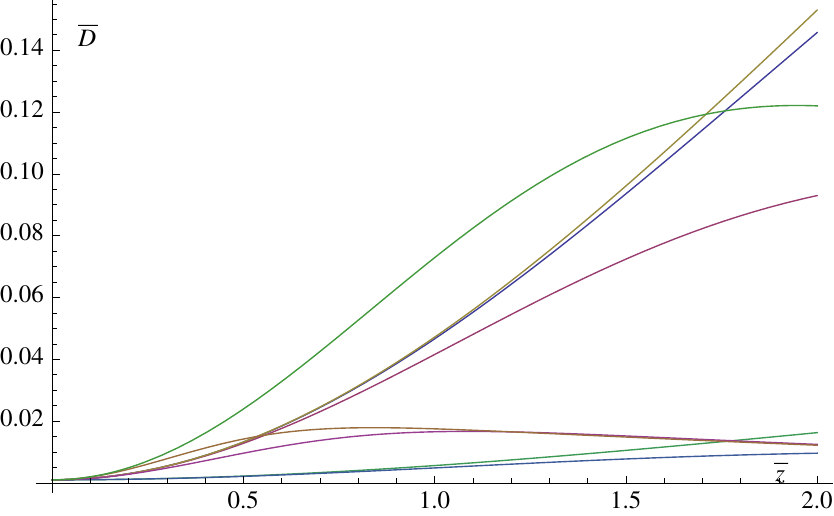} 
   \caption{Behaviour of $\hat{D}$ for several values of $\xi_{1}$ and $\xi_{2}$. The figure outlines an increasing behaviour of the function $\bar{D}$, and a maximum sometimes arises in the range of validity for the linear approximation.}
   \label{fig}
\end{figure}

Finally, the vertical equation, in the linear
approximation reads as
\begin{equation}
\partial _z\hat{p} +\omega ^2_K z\hat{\epsilon} = 0
\, ,
\label{vereq}
\end{equation}
which easily rewrites as an equation for
$\hat{\epsilon}$, i. e.
\begin{equation}
\frac{K_BT_0}{m_i}\partial _z\hat{\epsilon} +  \displaystyle\frac{\omega ^2_K}{2} z\hat{\epsilon}
+ \frac{3K_B\eta _0k_0k}{m_i\mathcal{N}c}\int Ydz = 0 
\, .
\label{verteq}
\end{equation}
Such an equation fixes the following form for
the perturbed mass density
$\hat{\epsilon} = \bar{\epsilon}(z^2)\sin k(r - r_0)$,
where the $z$-dependence comes from the ordinary
differential equation
\begin{equation}
\frac{K_BT_0}{m_i}\frac{d\bar{\epsilon}}{dz}
+ \frac{\omega ^2_K(r_0)}{2}z\bar{\epsilon}
+ \frac{3K_B\eta _0k_0k}{m_i\mathcal{N}c}
Y^0\int \exp \left\{-\frac{k_0z^2}{H_n}\right\}dz = 0
\, ,
\label{verteqod}
\end{equation}
having defined
$Y^0\equiv k_0\psi_1^0 /\partial _{r_0}\psi _0$.
The above equation, using the expression (\ref{formeta}) for $\eta _0$,
can be rewritten in the dimensionless form
\begin{equation}
\xi_{1}
\frac{d\bar{D}}{d\bar{z}}
+ \frac{\bar{z}}{2}\bar{D} 
+ \xi_{2}
   \mathrm{erf}\left(\frac{\bar{z}}{\sqrt{2}}\right) = 0
\, ,
\label{verteqodad}
\end{equation}
\begin{equation}\nonumber
\xi_{1}= \frac{K_BT_0}{m_i\omega ^2_K\Delta^2} > 1\,,\hspace{5mm}
\xi_{2} = \frac{2K_B\alpha {v_s}_0z_0 k_0k}{m_i\mathcal{N}c
\omega _K^2}Y^0\sqrt{\frac{\pi }{2}}
\end{equation}
where we defined $\bar{z}\equiv z/\Delta$,
$\bar{D}\equiv \bar{\epsilon}/\epsilon _0$ and erf is the error function: $\sqrt{\pi}$erf$(z)\equiv 2\int \exp(-z^{2})dz$.
This equation admits the formal solution
\begin{equation}\label{soluzione}
\bar{D} = e^{-\frac{\bar{z}^{2}}{4 \xi_{1}}}\left(\bar{D}(\bar{z}=0) - \xi_{2}\int_{0}^{\bar{z}}\displaystyle\frac{e^{\frac{u^{2}}{4 \xi_{1}}}\mathrm{erf}(\frac{u}{\sqrt{2}})}{\xi_{1}}du\right)
\end{equation}
which is plotted in fig \ref{fig} for several values of $\xi_{1}$ and $\xi_{2}$

We conclude this analysis,
by stressing that, because of (\ref{formeta}),
the expression (\ref{v-r}) provides the following
estimation for $v_r$
\begin{equation}
v_r\sim 2\alpha {v_s}_0k_0H_0\mathcal{O}(Y^0)
\, .
\label{estimat}
\end{equation}
We see that the accretion velocity is few orders of
magnitude less than the sound velocity on the equatorial
plane. In this respect, we observe that the parameter
$\alpha$ can be taken
in the range $10^{-3}-1$, while the dimensionless
term $k_0H_0\sim k_0H_n\sim \sqrt{\beta }$ can be
even much greater than unity and the linear approximation
requires $Y^0\ll 1$.

Despite the picture we fixed above is well-grounded
in the linear approximation,
it is well-known that the Nernst
coefficient is obtained in a kinetic theory
via an expansion in the inverse powers of the
cyclotronic frequency and it corresponds to a
first order approximation scheme.
Thus, this effect is, in principle, small
and we can expect that it be the proper explanation
to the accretion features of a crystalline disk,
in the weak field limit only, i.e. we have to require
$\beta > 1$, to ensure the model reliability.

\section{Concluding Remarks \label{secConclusioni}}

Our analysis was devoted to a study reconciling
the crystalline structure of the disk,
as outlined in \cite{C05}, with the existence
of an azimuthal equilibrium configuration.
In fact, the impossibility to neglect the radial
component of matter velocity, requested by the notion
of accretion, leads to include in the equilibrium
problem the angular momentum transport across the disk
and the unavoidable viscous features, associated to the
differential rotation.
The problem to get a self-consistent scheme, in which
the electron force balance equation properly holds
even along the azimuthal direction is then addressed.
In fact, the appearance of the radial and vertical
component of the matter velocity, together with the
vanishing nature of the azimuthal electric field in
an axisymmetric configuration, rise the question
how to get a consistent equilibrium balancing the
Lorenz force along the toroidal symmetry.
The proposal we investigate here concerns the role
that the Nernst coefficient can have in this perspective.
In agreement with the idea that this is a low cyclotronic
frequency effect, we treat only the linearised
MHD configuration.
The presence of such a collisional term
allows to link the radial and vertical velocity
components to the temperature gradients, as soon
as the azimuthal electron force balance equation stands.
The linear behaviour is completely self-consistent and
the adiabatic equation for the plasma pressure
is satisfied as a consequence of the incompressible
nature of the linear MHD, out-coming in this
axisymmetric scenario.
Finally we were able to establish the condition
ensuring that the radial temperature gradient
be negligible in the corresponding equilibrium,
while the linear vertical equilibrium is solved
exactly.
The main issue of our analysis is that the
radial velocity has an oscillating character, with an
amplitude few orders of magnitude below the 
sound velocity. Indeed, we deal with a non-zero
accretion rate because the material outside $r_0$ is
greater than the one inside (we are estimating the
matter density with its value at $r_0$). Such an oscillating profile
is a direct consequence of the linear crystalline structure
of the model. More than with a real matter in-fall,
arising expectably in the non-linear regime,
we get a matter flow which explains the formation of
the ring configuration within the disk.
However our goal opens interesting perspective
on the understanding about the
existence of a ring-like decomposition in
presence of a mechanism of accretion, i.e.
significantly non zero radial in-fall of the plasma.

\section{Acknowledgements}

We would like to thank B. Coppi for having attracted our attention on the relevance of the electron force balance in an accretion disk, F. Zonca for the interesting discussion on plasma physics in axis-symmetric configuration, and A. Corsi for her advices on the literature concerning accretion disk morphology. 

This work was developed within the framework of the CGW Collaboration
({\sf www.cgwcollaboration.it)}.
%\bibliography{bibliogio.bib}

\begin{thebibliography}{10}

\bibitem{C05}
B.~{Coppi}.
\newblock \emph{Physics of Plasmas} \textbf{12}(5), 057302 (2005).

\bibitem{verbunt82}
F.~{Verbunt}.
\newblock \emph{Space Science Reviews} \textbf{32}, 379 (1982).

\bibitem{S73}
N.~I. {Shakura}.
\newblock \emph{Soviet Astronomy} \textbf{16}, 756 (1973).

\bibitem{pringlerees72}
J.~E. {Pringle} and M.~J. {Rees}.
\newblock \emph{Astronomy \& Astrophysics} \textbf{21}, 1 (1972).

\bibitem{lyndenpringle74}
D.~{Lynden-Bell} and J.~E. {Pringle}.
\newblock \emph{Monthly Notices of the Royal Astronomical Society}
  \textbf{168}, 603 (1974).

\bibitem{rw}
R.~{Ruffini} and J.~R. {Wilson}.
\newblock \emph{Physical Review D} \textbf{12}, 2959 (1975).

\bibitem{CR06}
B.~{Coppi} and F.~{Rousseau}.
\newblock \emph{The Astrophysical Journal} \textbf{641}, 458 (2006).

\bibitem{B01}
G.~S. {Bisnovatyi-Kogan} and R.~V.~E. {Lovelace}.
\newblock \emph{New Astronomy Review} \textbf{45}, 663 (2001).

\bibitem{lynden-bell}
D.~{Lynden-Bell}.
\newblock \emph{Monthly Notices of the Royal Astronomical Society}
  \textbf{279}, 389 (1996).

\bibitem{cproceeding}
B.~{Coppi}.
\newblock \emph{Proceedings of the 2007 conference on Plasma Physics of the
  European Physical Society}  (2007).

\bibitem{SS88}
N.~I. {Shakura} and R.~A. {Sunyaev}.
\newblock \emph{Advances in Space Research} \textbf{8}, 135 (1988).

\bibitem{F37}
V.~C.~A. {Ferraro}.
\newblock \emph{Monthly Notices of the Royal Astronomical Society} \textbf{97},
  458 (1937).

\bibitem{c08}
B.~Coppi.
\newblock \emph{EPL} \textbf{82}(1), 19001 (2008).

\end{thebibliography}
%\bibliographystyle{gio}

%\end{document}

\end{document}